\documentstyle [aps,multicol,epsfig] {revtex}
\draft
\begin{document}
\def\mean#1{\langle#1\rangle}
\title {Multifractal characterization of stochastic resonance}

\author{ Alexander Silchenko~$^{1,2}$ and Chin-Kun Hu~$^1$}
\address{$^1$ Institute of Physics, Academia Sinica, Nankang, Taipei 11529,
Taiwan,\\ 
Saratov, Astrakhanskaya st.83, 410026 Russia.
$^2$ Department of Physics, Saratov State University,
}
\date{\today}
\maketitle
\begin{abstract}
We use a multifractal formalism to study the effect of stochastic 
resonance in a noisy bistable system
driven by various input signals.
To characterize the response of a stochastic bistable system we introduce a new
measure based on the calculation of a singularity spectrum for a  return time 
sequence. We use wavelet transform modulus maxima method for the 
singularity spectrum computations.
It is shown that the degree of multifractality defined as a width of 
singularity spectrum can be successfully used as a measure of complexity 
both in the case of periodic and 
aperiodic (stochastic or chaotic) input signals.
We show that in the case of periodic driving force singularity 
spectrum can change its structure qualitatively becoming  monofractal in the 
regime of stochastic synchronization. This fact allows us to consider the 
degree of multifractality as a new measure of stochastic synchronization also.
Moreover, our calculations have shown that the effect of stochastic resonance 
can be catched by this measure even from a very short return time sequence. 
We use also the proposed approach to characterize the  
noise-enhanced dynamics of a coupled stochastic neurons model. 
\end{abstract}

\begin{multicols}{2}

\section{introduction}
It is well known that noise is present inevitably in  \,  all real processes. 
The reckoning of its influence is very important for a deeper understanding 
of the dynamics of real systems . Stochastic resonance (SR) discovered by 
Benzi, {\em et al.} and Nicolis {\em et al.} \cite{benzi-nicolis} during the
study of the Ice Ages 
is one of the bright examples of a nontrivial noise action on a nonlinear 
system. As a model of climate dynamics they  proposed to consider 
a bistable system simultaneously driven by noise and a periodic signal. 
It was shown that  tuning the noise level, an enhancement of 
a bistable system's response to the  periodic force becomes possible.
Beginning from the late 1980s, a wealth of theoretical and experimental 
papers followed, extending the notion of SR 
(for extensive reviews, see \cite{jung,gammaitoni,anishchenko}) and 
discovering new applications in different fields of sciences.
There are a few different approaches to quantitative description  of 
stochastic resonance depending on the amplitude and character of an input 
signal which may be periodic, chaotic or even  stochastic.\, 
Originally, the enhancement of a weak periodic input signal was characterized 
by the response amplitude at the frequency of periodic signal. 
Fauve and Heslot \cite{fauve} and McNamara, Wiesenfeld, and Roy \cite{roy}
suggested to use the signal-to-noise ratio (SNR) as a quantitative measure 
of SR. Both quantities, the amplitude and the SNR, undergo a resonance-like 
curve as a function of noise intensity. Spectrum power amplification 
defined in \cite{jung,hanggi} as the ratio of periodic components in output 
and input power spectrums demonstrates similar behavior taking a maximal 
value at an optimal noise level.

Other measures based on the residence-time distribution  were introduced for 
description of SR  in \cite{gam2,zhou}.
In this case the main object of considerations is the structure of the  
mentioned distribution that contains a series of peaks at the odd multiples
of the half-period of driving. All of them go through maxima as a function
of the noise strength \cite{zhou}. Gammaitoni {\em et al.} 
introduced the area under the peak of the residence-time distribution
at the half-period of driving as a measure of SR \cite{gam3}. 
Some later, a fully systematic theory for the residence-time distribution 
functions was developed by Choi, Fox and Jung  \cite{choi}.
They showed that to characterize correctly SR based on the
residence-time distribution it is necessary to find the 
difference between the residence-time distribution in the 
presence of the modulation and the residence-time distribution in the absence 
of the modulation at the half-period of the external force.
In \cite{bulsara} the receiver operating characteristic was used for
quantitative characterization of a response of coupled overdamped 
nonlinear dynamic elements driven by a weak sinusoidal signal embedded in 
Gaussian white noise. This approach was complemented and 
generalized in \cite{robinson} where SR was described in terms of 
maximization of information-theoretic distance measures 
between probability distributions of the output variable.

Evidently, the measures mentioned above can be used in the case when the 
input  signal has a clear distinguishable peak in its power spectrum. To 
characterize a response of a noisy nonlinear system on an aperiodic
driving force it is necessary  to use other measures. In order to 
estimate the response of a noisy excitable (or bistable) system to a weak 
aperiodic signal Collins {\em et al.} \cite{collins} introduced the 
input-output cross-correlation measures and a measure of transinformation 
quantifying the rate of information transfer from stimulus to response.
They showed, in particular, that the rate of information transfer between 
system output and input is optimized by noise and coined the term 
aperiodic stochastic resonance to describe this phenomenon.
The coherence function was used in \cite{shura1} to characterize the response 
of a bistable system to a weak stochastic input. It was calculated
analytically in the framework of the linear response theory (LRT) that has 
been successfully applied to SR and related phenomena \cite{jung,dykman}. 

From the practical point of view, it is very important to have 
the measures calculated from a sequence of the time intervals 
characterizing the dynamics of an object under study.  
Two examples of such sequences, which may be most popular at present,
are interbeat interval time series from cardiophysiology 
and neurons spike trains from neurodynamics. 
The statistical analysis of heart-beat dynamics have shown that the 
use of approaches basing on wavelet or Hilbert transform (or their dual use) 
has a number of benefits in comparison with the traditional ones such as 
power spectrum and correlation analysis \cite{ivanov1}. 
It allows to analyze the information stored in the Fourier phases of a 
signal under study which is crucial for determination of nonlinear 
characteristics. As was recently discovered by Ivanov 
{\em et al.} \cite{ivanov2} the human heartbeat dynamics possesses the 
multifractal properties. They used the wavelet-based approach 
developed in \cite{halsey,arneodo,hwang,muzy1,muzy2,bacry,muzy3} to the 
analysis of complex non-stationary time series.  
It has been shown that a heartbeat sequence of a healthy subject has a 
multifractal scaling whereas the data from subjects with a pathological 
condition demonstrate a loss of multifractality.
Moreover, authors demonstrated an explicit relation between the nonlinear 
features (represented by the Fourier phase interactions) and the 
multifractality of healthy cardiac dynamics.
The efficiency of their approach based on the time-frequency 
localization properties of wavelets allowing to analyze the 
non-stationarityies in time series.

It is reasonable to try to use the same approach for the quantitative 
characterization of SR when a  sequence under 
study is defeated by the concerted action of an external signal and of a 
random force. In this case, an input signal (periodic or aperiodic) is the 
source of non-stationarity in a response which can be analyzed by means of 
the  wavelet-based algorithm mentioned above. From this point of view, 
it is naturally to expect that the use of the multifractal formalism will 
allow us to introduce a new universal measure quantifying SR for an arbitrary 
external signal.

The main goal of the present study is the description of SR  from the 
multifractal analysis point of view. Basically, we treat as a model a
bistable system simultaneously driven by the white noise and an input signal.
To characterize the scaling properties of a return time sequence we use 
a spectrum of local H\"older exponents calculated by means of the
wavelet transform modulus-maxima (WTMM) method.
For this purpose, we first give some necessary definitions and illustrate  
the procedure of singularity spectrum calculation in Sec.~II.
Section~III deals with the multifractal analysis of the stochastic bistable 
system response for different kinds of  input signals. In this section,
we introduce a new measure for quantitative description of SR  and 
stochastic synchronization. We also test it 
ability to catch SR for different lengths of a return time sequence.
In Sec.~IV we apply mentioned approach to the  study of aperiodic 
stochastic resonance and coherence resonance in an unidirectionally 
coupled neurons model. In Sec.~V we summarize our results and discuss
the advantages of our measure in comparison with traditional measures.

\section{singularity spectrum and wavelet-transform modulus-maxima 
method}

It is well known that stochastic signals can be conditionally divided into 
two different classes.  The first one includes the homogeneous 
signals characterizing by a single global 
Hurst exponent and having the same scaling properties at all time intervals.
The second one includes the multifractal signals to describe the scaling properties
of which it is necessary to use many local Hurst exponents (or H\"older 
exponents) quantifying the local singular behavior and local scaling 
in time series. According to the definition \cite{muzy3}, the H\"older 
exponent $h(x_0)$ of a function $f$ at the point $x_0$ is the greatest 
$h$ so that $f$ is Lipschitz at $x_0$, i.e., there exists a constant $C$ 
and a polynomial $P_n (x)$ of order $n$ so that for all $x$ in a neiborhood 
of $x_0$ we have 
\[
\mid f(x)-P_{n}(x-x_0)\mid\,   \le C\mid x-x_0\mid^h.
\]
In fact, it  measures the degree of irregularity of $f$ at the point $x_0$.
The singularity spectrum $D(h)$ of the signal can be defined as the function 
that gives for a fixed $h$, the Hausdorff dimension of the set of points $x$ 
where the exponent $h(x)$ is equal to $h$.

As was mentioned above, to determine the whole singularity spectrum $D(h)$ 
from an experimental signal it is necessary to use the approach based on the 
wavelet transform (WT), which permits an analysis both in physical space and 
in scale space. The WT of the function $f$ is defined as 
\begin{equation}
\label{wt}
T_{\psi}(b,a)=\frac{1}{a}\int_{-\infty}^{+\infty} \psi \left(\frac{x-b}{a} 
\right)\, f(x)\, dx,
\end{equation}
where $\psi$ is the analyzing wavelet, $a \in R^{+}$ is a scale parameter and 
$b \in R$ is a space parameter. The analyzing wavelet $\psi$ is generally 
chosen to be well localized in both space and frequency domain.
A class of widely used real-valued analyzing wavelets which satisfies the 
above condition is given by the Gaussian function and its derivatives.
As was proven by Mallat and Hwang \cite{hwang} the WT modulus maxima (local 
maxima of $\mid T_{\psi}(x,a)\mid$ at a given scale $a$) detect all the 
singularities of a signal under study.
The skeleton from the modulus maxima lines contains all the information about 
the hierarchical distribution of singularities in the signal.
The WTMM method consists in taking advantage of the space-scale partitioning 
given by this skeleton to define a partition function which scales, in the 
limit $a \rightarrow 0^{+}$, in the following way \cite{arneodo,muzy2,bacry}:
\begin{equation}
\label{pf}
Z(q,a)= \sum_{\{x_i (a)\}_i} \,\mid 
T_{\psi}(x_i (a),a)\mid ^q\,  \sim \, a^{\tau(q)},
\end{equation}
where $\{x_i (a)\}_i$ are the WT modulus maxima and $q\in R$. According to 
the theorem proved in \cite{bacry}, $D(h)$, 
the singularity spectrum of the function $f$, is obtained by Legendre 
transformation of the function $\tau(q)$ defined in (\ref{pf})
\[
D(h)=\min_q (qh-\tau (q)).
\]
The variables $h$ and $D(h)$ play the same 
role as the energy and entropy in the thermodynamics, whereas instead of 
the inverse of temperature and  free energy we have $q$ and $\tau (q)$ 
\cite{muzy2,muzy3}.
From a numerical point of view, it is more
conveniently to calculate at first the scaling exponents:
\[
%\begin{equation}
h(q)=\lim_{a\rightarrow 0}\frac{1}{\ln a}\sum_{\{x_i (a)\}_i}\!
\tilde{T}_{\psi}(q;x_i (a),a)\ln \mid T_{\psi}(x_i (a),a)\mid,
%\end{equation}
\]
and
\[
D(h(q))=\lim_{a\rightarrow 0}\frac{1}{\ln a}\sum_{\{x_i (a)\}_i}\!
\tilde{T}_{\psi}(q;x_i (a),a)\ln \tilde{T}_{\psi}(q;x_i (a),a),
\]
where $\tilde{T}_{\psi}(q;x_i (a),a)=\mid T_{\psi}(x_i,a) \mid^q\!/
\sum_{x_i}\!\mid T_{\psi}(x_i,a) \mid^q$.
Further, we extract the set of H\"older exponents and 
corresponding singularity spectrum $D(h)$ from log-log plots of $h(q)$ and 
$D(h(q))$ \cite{muzy1}.

As a simple example, we calculated $\tau(q)$ and $D(h)$ for 
the ordinary Brownian motion which is characterized by the single 
global Hurst's exponent $H=1/2$.
Figure 1  demonstrates clearly the homogeneous scaling for the 
ordinary Brownian motion.
For more detailed information about calculation's procedure and 
additional references on freely distributed software see \cite{mallat}. 
\begin{figure}
\epsfig{figure=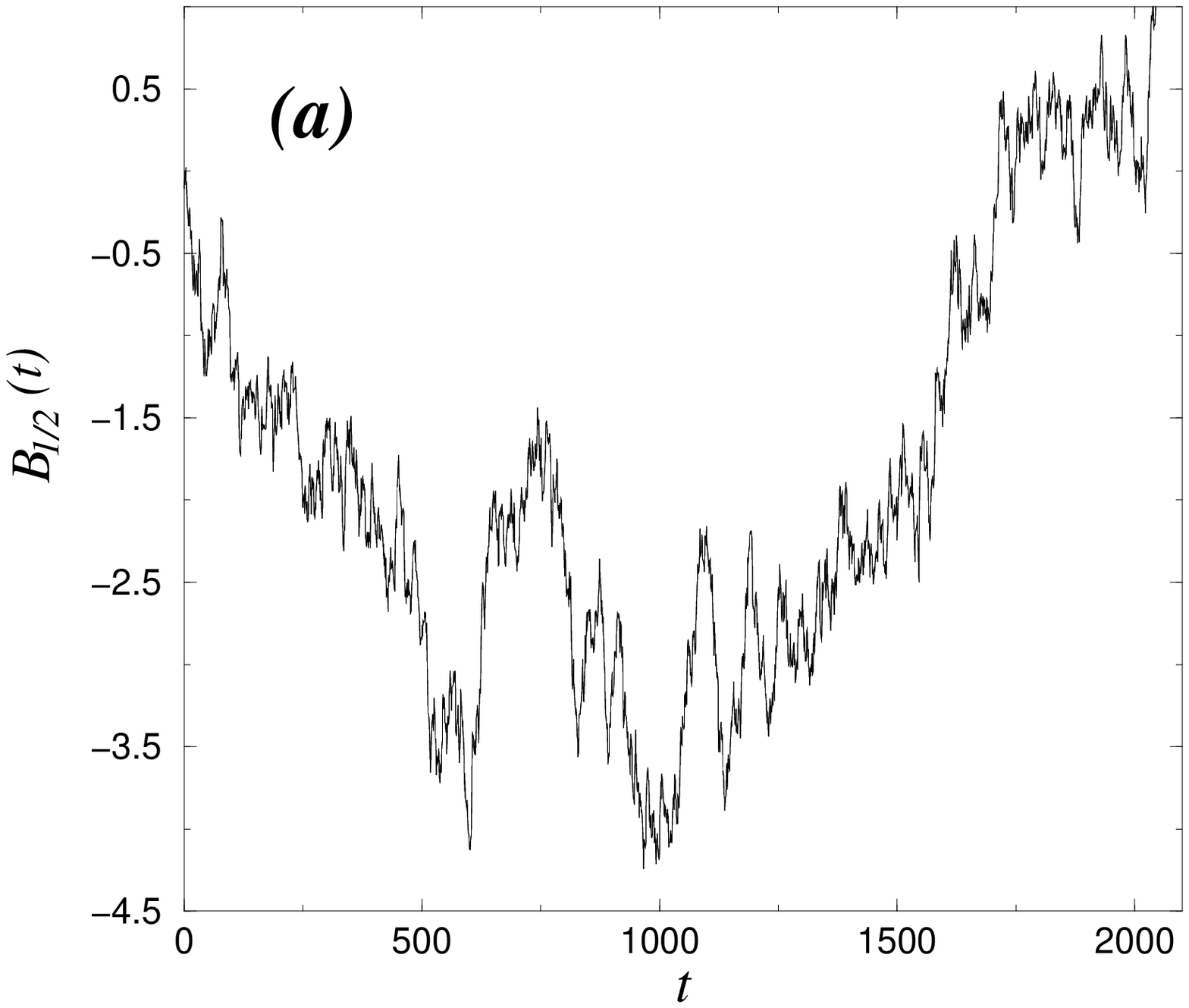,width=7.5cm,height=3.7cm}\\
\vskip 0.05cm
\hskip 0.3cm
\epsfig{figure=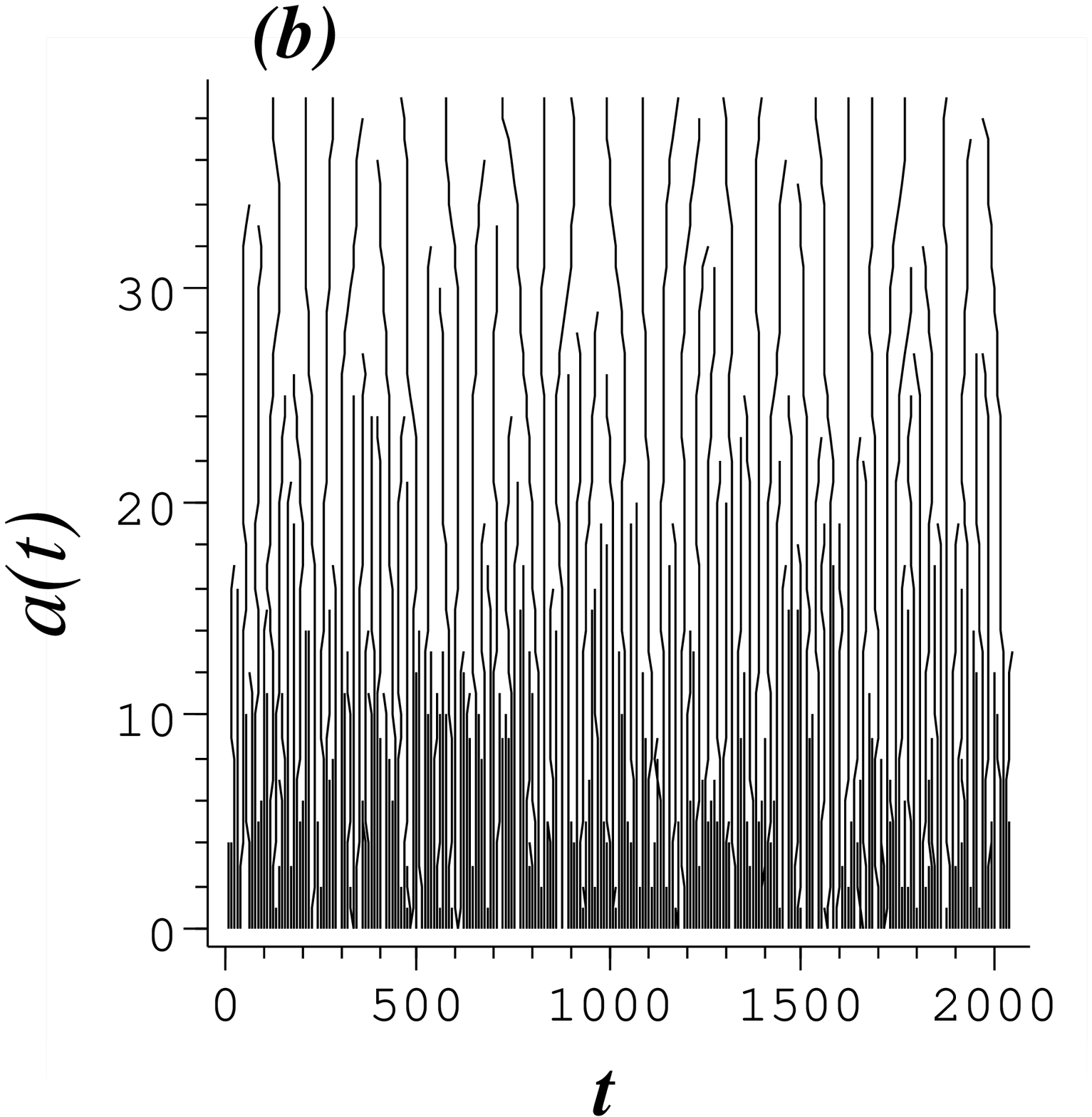,width=7.4cm,height=3.7cm}\\
\vskip 0.2cm
\hskip -0.2cm
\epsfig{figure=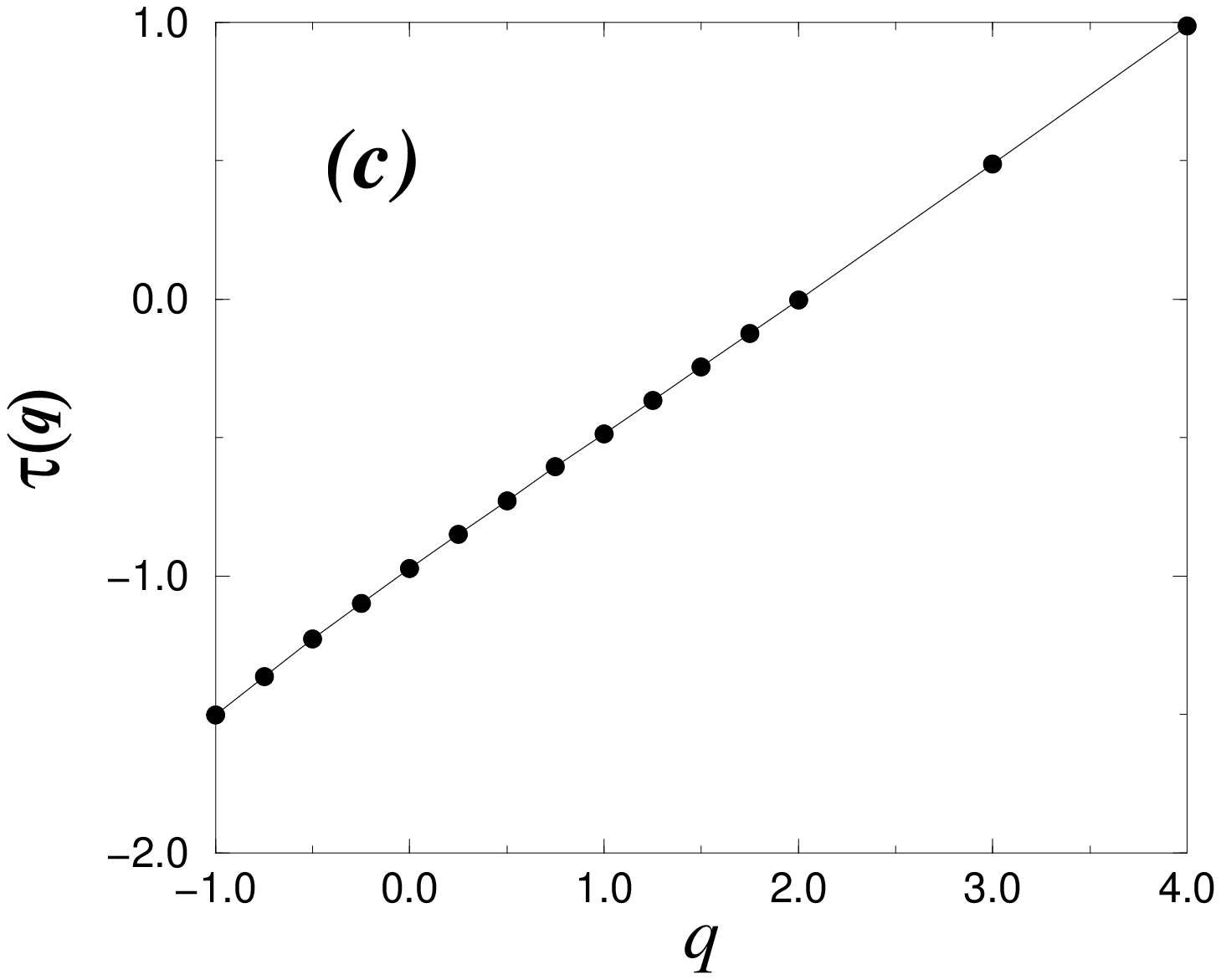,width=4.2cm,height=3.7cm}
\epsfig{figure=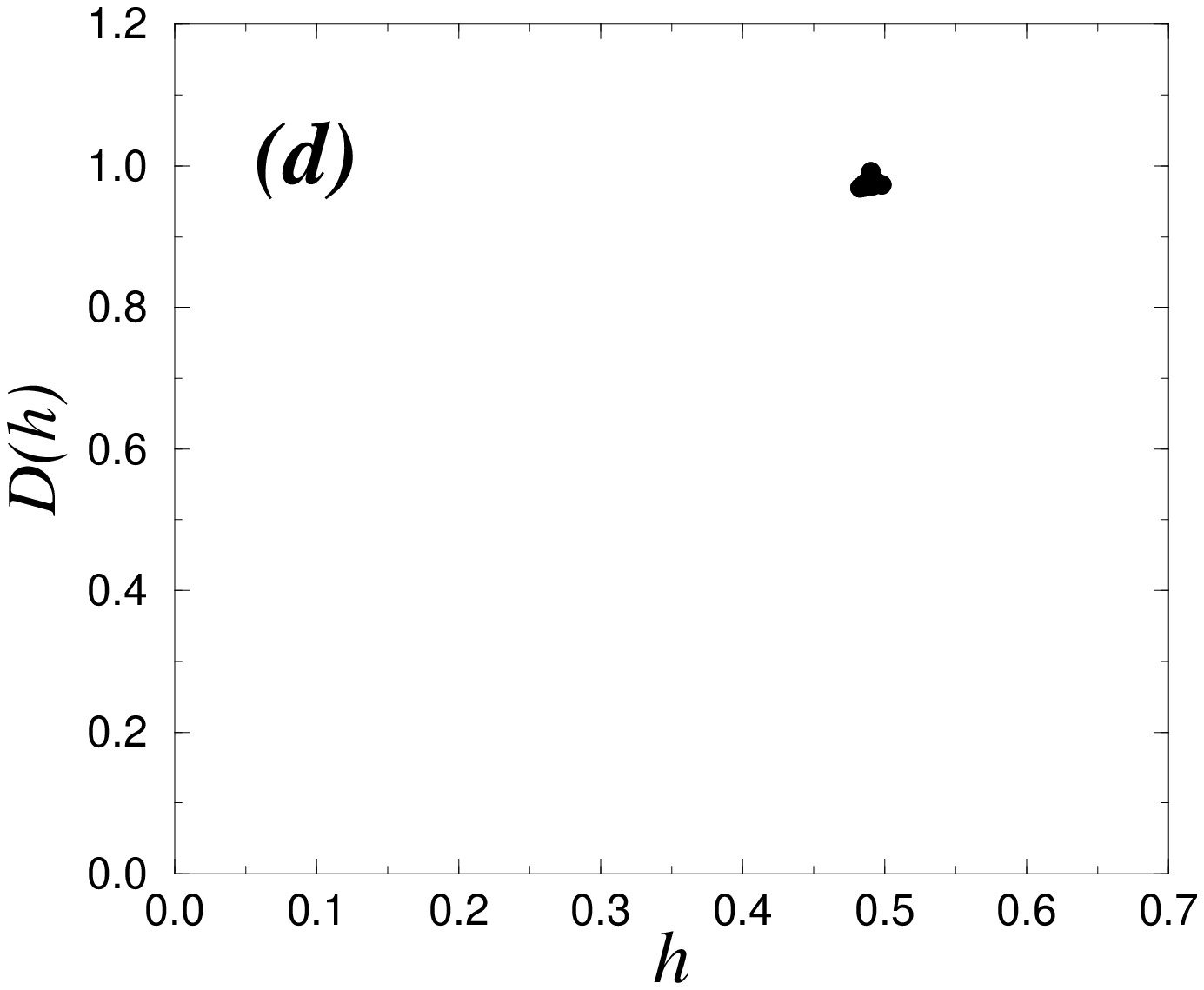,width=4.2cm,height=3.6cm}
\vskip 0.2cm
\begin{minipage}{8.4cm}
\caption{(a) -- the path of a Brownian particle; (b) -- the modulus-maxima 
skeleton of the random signal pictured in (a); 
(c) -- the dependence $\tau (q)$; (d) -- the  singularity spectrum. 
The first derivative of the Gaussian function was used as the analyzing wavelet.}
\label{classic}
\end{minipage}
\end{figure}

\section{multifractal approach to stochastic resonance}
We treat for our study the overdamped bistable 
oscillator which is governed in canonical units \cite{jung} by the following 
stochastic differential equation (SDE):
\begin{equation}
\dot{x}= x - x^{3} + \sqrt{2Q}\,\xi(t)+ y(t), 
\label{model}
\end{equation}
where $\xi (t)$ is the white Gaussian noise with the correlation
function $\mean{\xi (t)\: \xi (t')}\: =\: \delta (t-t')$ and
$\mean{\xi (t)}\: =\: 0$, $y(t)$ is an input signal.

In the following subsections, we present the results of numerical simulations 
of Eq.~(\ref{model}) for different kinds of input signal.
To characterize SR from the multifractal formalism point of view we will 
process the sequences of the return times to the one of the potential wells  
\cite{jung} normalized on a characteristic time scale of an external 
driving force.

\subsection{Periodic input signal}
Let us start from the more  simple and  well studied case of periodic 
external driving when $y(t)=A\,\sin(\Omega t +\phi)$.
The amplitude of the input signal is assumed to be small, i.e., the signal 
alone cannot switch the system from one state to another in the absence of 
noise. For the low-frequency periodic modulation 
considered in this paper this means 
\begin{equation}
A\le A_0 =\frac{2}{3\sqrt{3}}.
\end{equation} 
As is well known, the study of SR can be conditionally separated on the two
cases. The first one deals with the situation of a weak input signal when the 
amplitude of periodic driving is very small in comparison with a potential 
barrier and can be considered in the framework of the LRT \cite{dykman,shura1}.
The second one is beyond the limits of LRT and corresponds to an 
amplitude of a subthreshold input signal comparable with a barrier. 
In the last case, the dynamics of a bistable system is characterized by 
a high degree of coherence between the  switching process and input 
signal \cite{boris}.
It can be correctly described in terms of the phase synchronization theory both
for periodic \cite{shurame}, chaotic \cite{me} and stochastic 
input signals \cite{shura2}. 
Moreover, as was shown in \cite{shura3} SR takes the form of 
the noise-induced order in this case. Such information-theoretical 
measures as the source entropy and the dynamical entropy display a minimum 
at an optimal noise intensity. 
\begin{figure}
\epsfig{figure=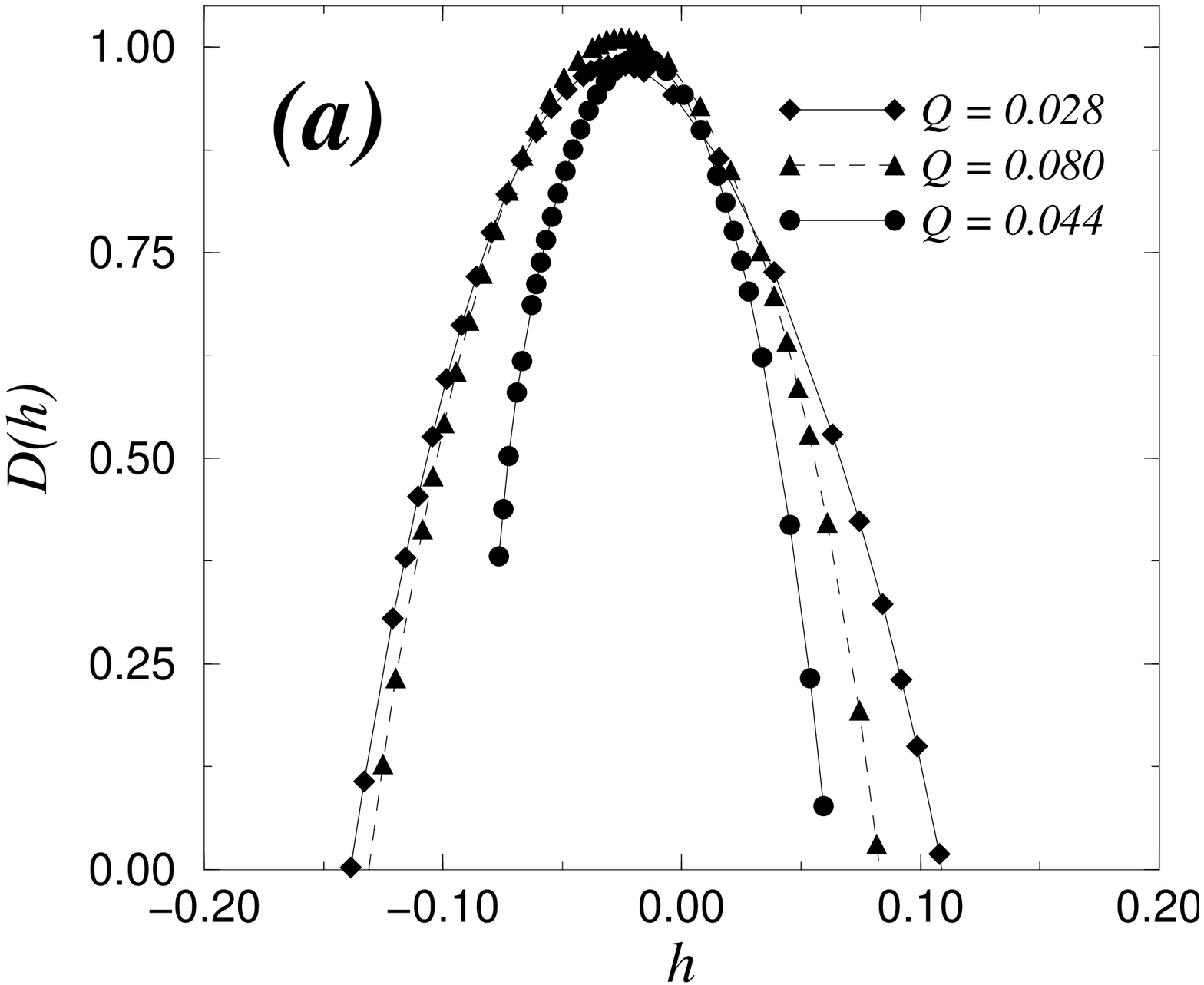,width=4.3cm}
\hskip 0.1cm
\epsfig{figure=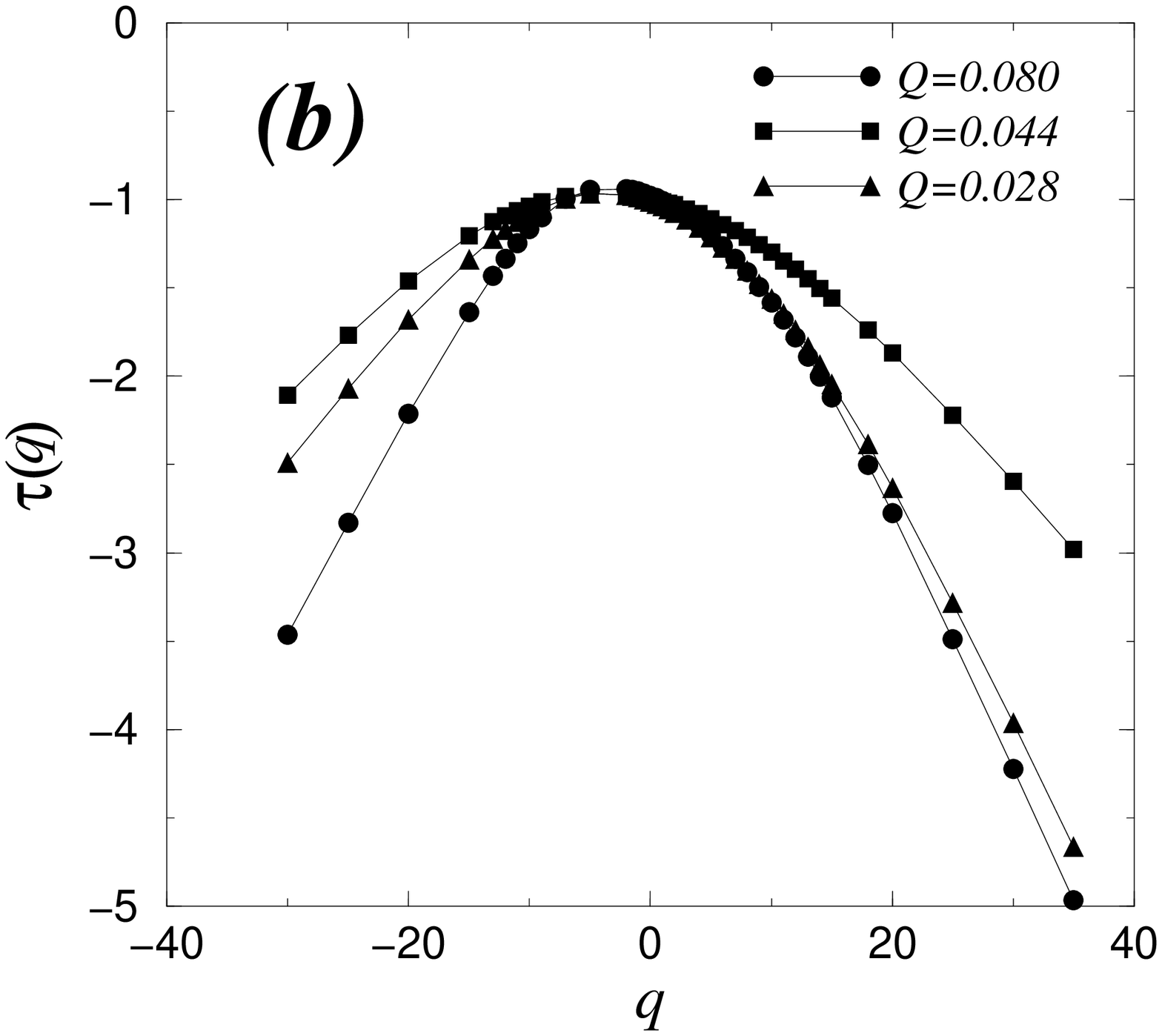,width=4.2cm}
\vskip 0.1cm
\epsfig{figure=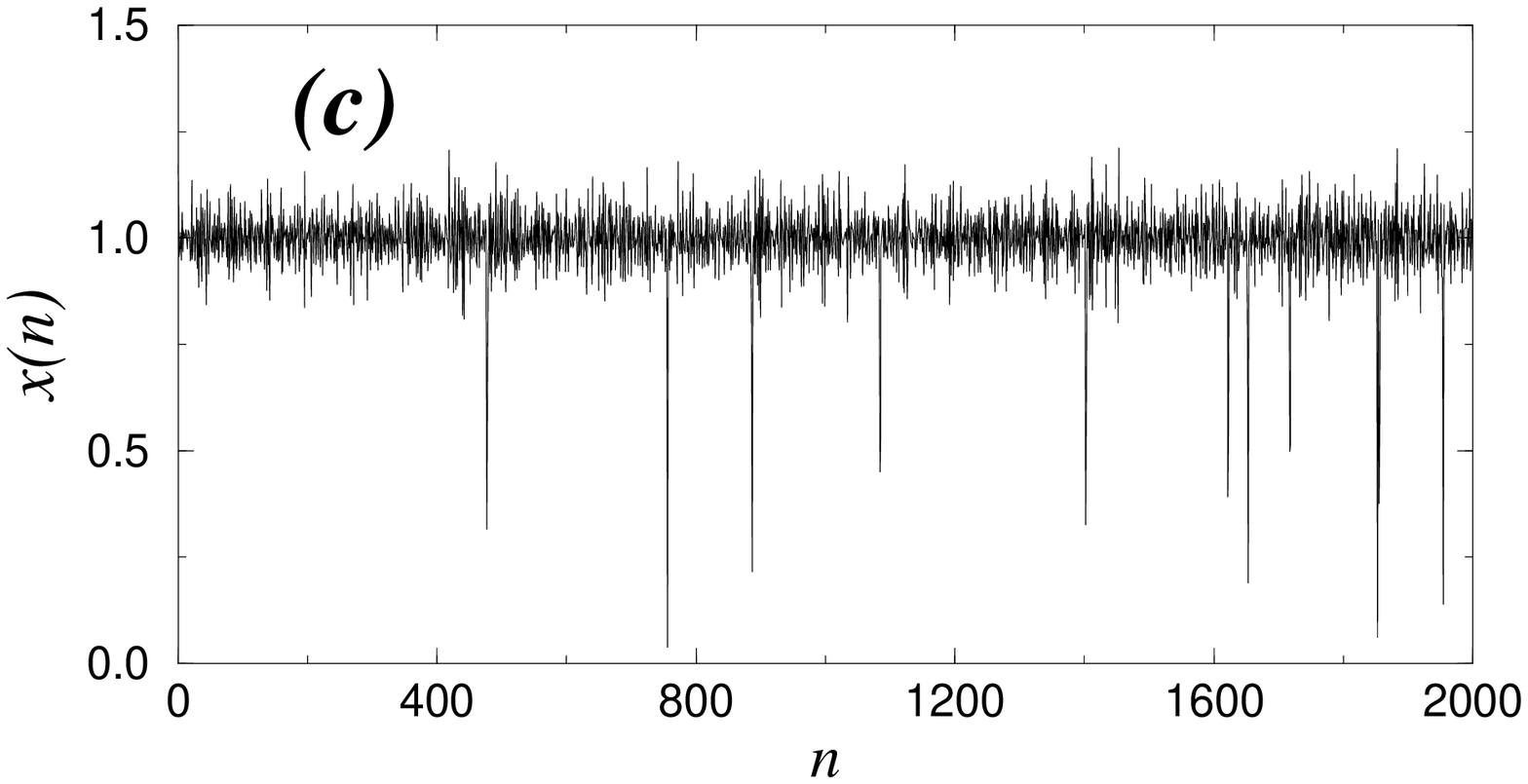,width=4.25cm,height=3cm}
\hskip 0.1cm
\epsfig{figure=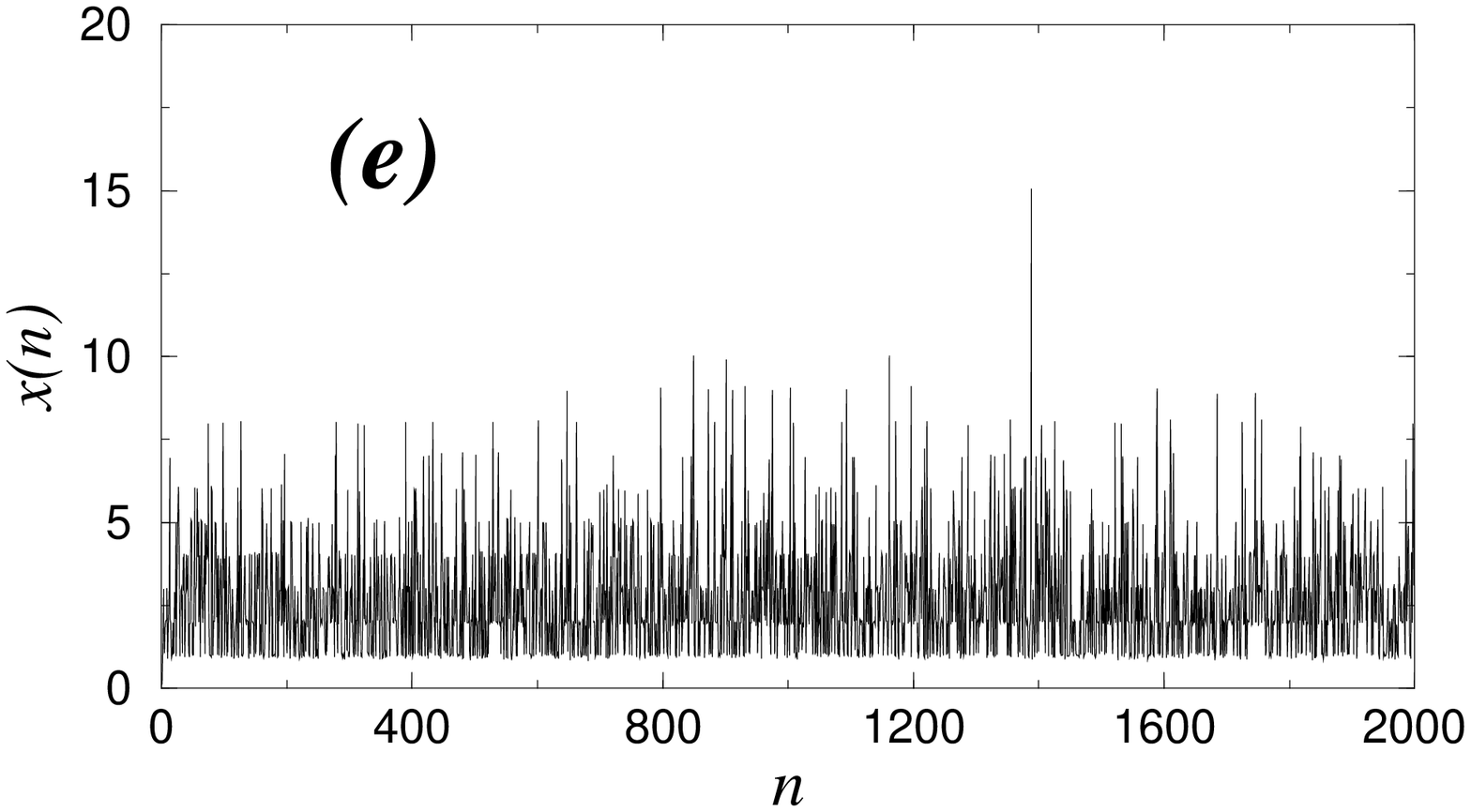,width=4.25cm,height=3cm}
\vskip 0.1cm
\centering{
\epsfig{figure=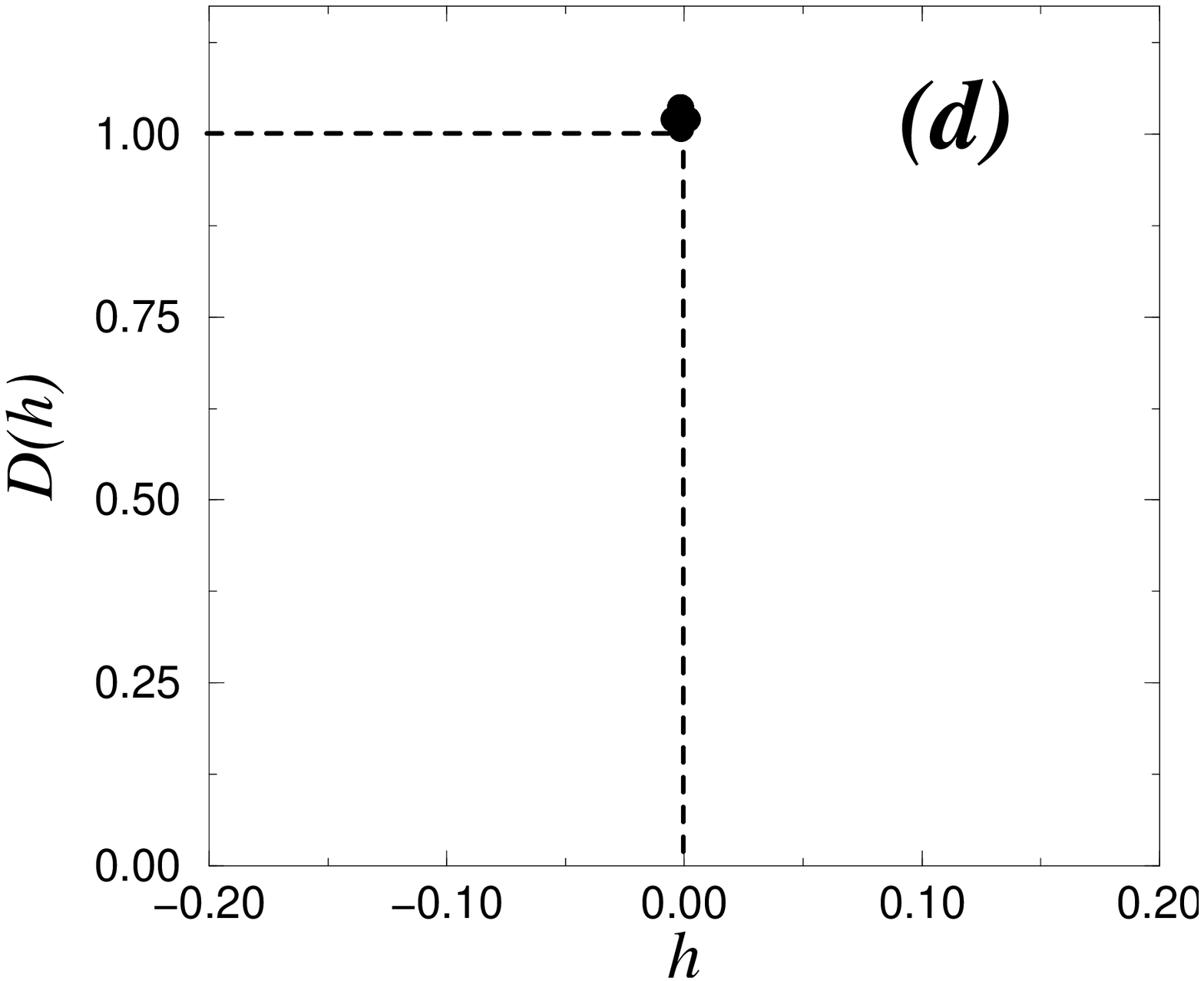,width=4.2cm}
\hskip 0.1cm
\epsfig{figure=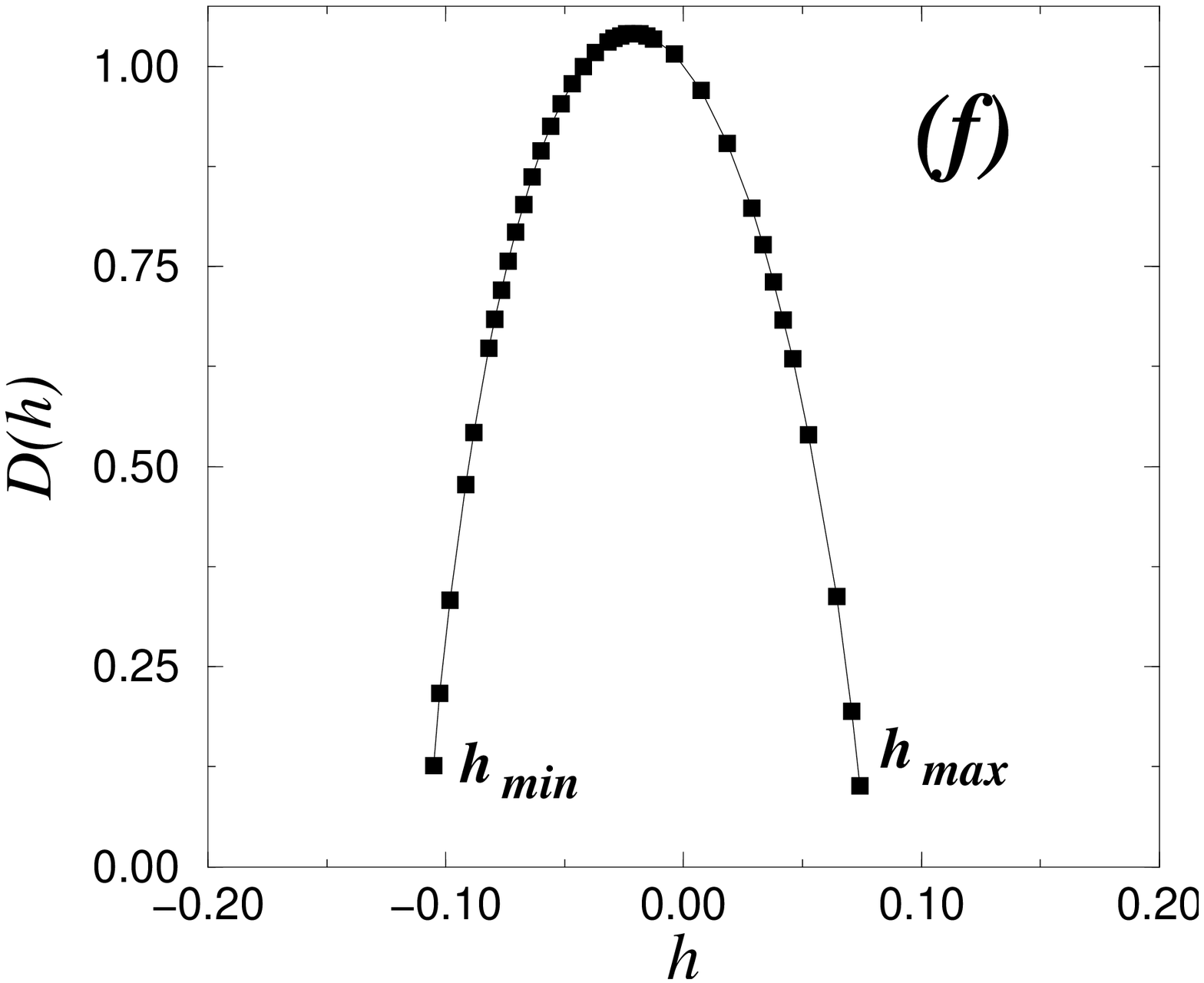,width=4.2cm}
}
\vskip 0.1cm
\begin{minipage}{8.4cm}
\caption{(a) $D(h)$ in the case of the weak periodic 
input signal $A\: =\: 0.081, \Omega\: =\: 0.004$ for different values of 
noise intensity $Q$; 
(b) $\tau(q)$ corresponding to the singularity spectrums, $D(h)$, 
in (a) for different values of noise intensity; 
the return time sequences normalized on the external force period and 
corresponding singularity spectrums in the case of $A\: =\: 0.286, 
\Omega\: =\: 0.004$ for $Q=0.040$ ((c) and (d)) and $Q=0.012$ ((e) and (f)).
The Gaussian function was used as the  analyzing wavelet.}
\label{periodic}
\end{minipage}
\end{figure}
\noindent
In order to describe quantitatively these different situations from 
the multifractal formalism point of view, we calculated singularity spectrums 
for different values of the driving amplitude taking as the signal under study  
a sequence of the return times normalized on the driving period. Each sequence 
contained 10000 points.
The results of calculations have shown the following. Singularity spectrum 
of the response has a bell shape form both in the presence and in the 
absence of the external periodic driving. That caused by a nonlinearity of the 
bistable system's response to the external driving force which manifests itself in 
the presence of the different modes and in their interaction.
As seen from  Fig.~2, the tuning of noise level in the 
system of Eq. (\ref{model}) leads to the changes 
in singularity spectrum both for 
weak and for strong enough driving signals. The width of the singularity 
spectrum takes its minimal value for an optimal noise intensity. 
It remains finite in the case of a weak periodic driving force, whereas in 
the regime of stochastic synchronization singularity spectrum qualitatively 
changes its form shrinking to a single point. 
The return times fluctuate around the driving period in the regime of 
noise-enhanced phase locking and large bursts are seldom happen 
(Fig.~2 $(c)$),
while for the values of noise 
intensities lying outside of synchronization region the respectively large 
fluctuations dominate (Fig.~2 $(e)$). 
\begin{figure}
\epsfig{figure=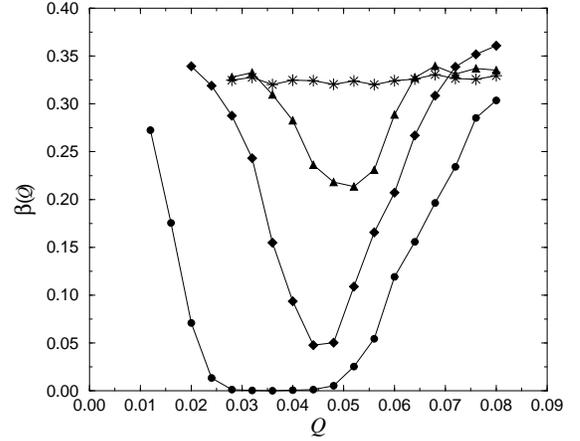,width=7.5cm}
\vskip 0.05cm
\begin{minipage}{8.2cm}
\caption{The degree of multifractality, $\beta$, vs. noise intensity 
for different values of the periodic force 
amplitude: $A\: =\: 0$ (stars), $A\: =\: 0.081$
(triangles), $A\: =\: 0.152$ (diamonds), $A\: =\: 0.268$ (circles).
The frequency of external force has the same value as in 
Fig.~\ref{periodic}
The Gaussian function was used as the  analyzing wavelet.}
\label{beta-periodic}
\end{minipage}
\end{figure}
It is natural to consider the degree of multifractality $\beta$
defined as the difference between the maximal and minimal H\"older exponents 
belonging to the one and the same singularity spectrum as a new measure for 
SR and stochastic synchronization.
As clearly seen from Fig.~3, the dependence of $\beta$ 
on the noise intensity is characterized by the presence of a minimum both for
weak and for sufficiently strong input signals.
In the regime of stochastic synchronization $\beta$ equals to zero that 
corresponds to a single point singularity spectrum (see Fig.~2 (d)). 
In this case the scaling features of the return time sequence under study 
is characterized by the single scaling exponent that caused  by the 
linearization of the response in the regime of switchings synchronization.
Indeed, the interaction between different modes in response is suppressed in 
synchronous regime, because the switchings in the system (\ref{model}) are 
in-phase with the input signal. The mode corresponding to the periodic input 
signal dominates and suppresses all others. 
\begin{figure}
\epsfig{figure=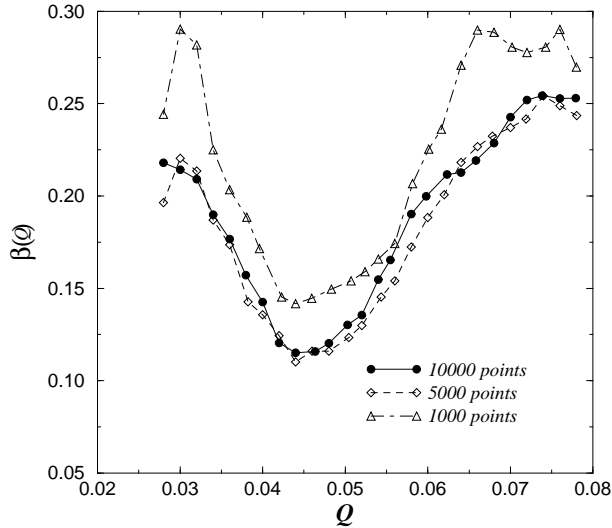,width=8.2cm}
\vskip 0.2cm
\begin{minipage}{8.2cm}
\caption{ Regions of stochastic synchronization constructed by means of 
effective diffusion constant (dashed line) and of multifractality degree
(solid line). Driving frequency is $\Omega \:=\: 0.004$. The Gaussian
function was used as the analyzing wavelet.}
\end{minipage}
\end{figure}
\noindent
In turn, it leads to 
simplification of the singularity spectrum that reflects the absence of 
interactions between Fourier phases in response.
It is necessary to emphasize here, that we understand stochastic 
synchronization as  instantaneous matching of the input/output phases which is 
observed in a finite region on the parameter plane 
``noise intensity -- amplitude of periodic force''.
Traditionally, synchronization in noisy nonlinear 
oscillatory systems is estimated quantitatively by means of effective 
diffusion constant  which characterizes a velocity of spreading of an 
initial phase difference distribution \cite{stratonovich}.
Recently, this classical approach to synchronization was successfully 
used for quantitative description of the noise-enhanced phase coherence 
which takes place in stochastic bistable system driven by a
subthreshold external signal \cite{shurame,me,shura2}. 
The effective diffusion constant demonstrates a minimum decreasing up to
a very small value in the region of synchronization. 
The above introduced wavelet-based measure demonstrates exactly the same 
behavior taking the zero value in the phase-locking regime that allows us to 
consider it as the measure of stochastic synchronization as well.
\begin{figure}
\epsfig{figure=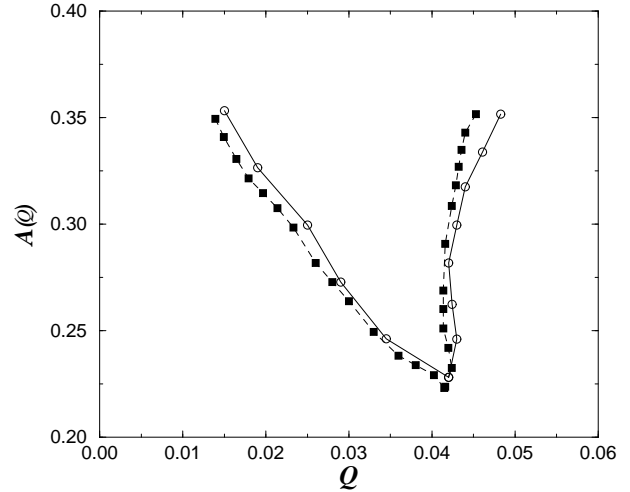,width=8.2cm}
\vskip 0.2cm
\begin{minipage}{8.2cm}
\caption{The degree of multifractality vs. noise intensity calculated for 
indicated values of the length of the return time sequence. 
The amplitude and frequency of the driving force are 
$A\: =\: 0.12$, $\Omega\: =\: 0.004$.}
\end{minipage}
\end{figure}
Using $\beta$ as the measure of stochastic synchronization it is 
possible to construct the region of synchronization on the parameter plane 
``noise intensity - amplitude of periodic force'' (see Fig.~4).
Inside of this region singularity spectrum of the response remains monofractal 
that once more time demonstrates the simplification of the response in the 
regime of synchronization. 
Synchronization region has a tongue-like form and nearly coincides with the 
similar one constructed by means of the effective diffusion constant used in 
\cite{shurame,me,shura2}. It should be noted that the degree of 
multifractality does not relate directly with effective diffusion 
constant. The multifractal approach based on the analysis of the scaling 
features of the temporal sequence whereas the effective diffusion constant 
is the value characterizing a probability distribution of the input/output 
instantaneous phase difference.
Our numerical studies also have shown that  $\beta$ demonstrates a 
monotonous  dependence on the driving frequency.

The length of the analyzed signals becomes the important parameter 
if they obtained in real experiments with live objects. 
Evidently, to estimate the possibility to use the above proposed measure 
in real situations we need to test its ability to catch SR for different 
lengths of the return time sequences. The results of our computations have 
been shown that the degree of multifractality takes the possibility to 
observe SR both for  long and for sufficiently short return time sequences. 
As seen from Fig.~5, $\beta$ calculated over the short return time sequences 
containing only 1000 points demonstrates nearly the same behavior as in the 
case of the long sequences.
\begin{figure}
\epsfig{figure=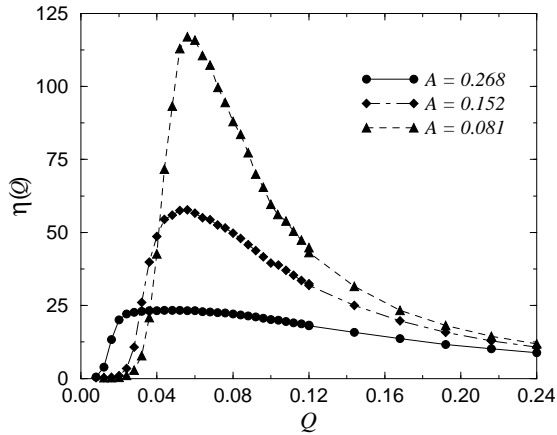,width=7.5cm}
\vskip 0.2cm
\begin{minipage}{8.2cm}
\caption{Spectral power amplification vs. noise intensity for different
values of the driving amplitude. Driving frequency is $\Omega \:=\: 0.004$.}
\end{minipage}
\end{figure}

Thus, the width of singularity spectrum manifests itself as the general and 
universal characteristic for description of SR, because 
it works very well in a wide range of values of 
the driving amplitude and frequency. Its calculation allows us to get the 
information about the scaling properties of the frequency fluctuations 
of a nonlinear system response and doesn't requires any information 
about input signal. Moreover, that new characteristic allows one to analyze 
effectively a response even in the case of short return time sequences 
that has an especial meaning for possible applications.

The obtained results are in a good agreement with the behavior of the spectral 
power amplification $\eta$ firstly used for quantitative characterization 
of SR in \cite{jung,hanggi}. It demonstrates a maximum the absolute value of 
which is increased with the decrease of the input driving 
amplitude (see Fig.~6).
The values of noise intensity maximizing $\eta$ nearly coincide with the ones
minimizing  the degree of multifractality. The growth of the driving amplitude 
leads to the shift of minimal values of $\beta$ to lower noise level 
as well as for spectral power amplification.

\subsection{Stochastic input signal}

From the practical application point of view, it is  more interesting 
the situation when an input signal has a complex structure.
SR for the input signals with fluctuating amplitude and phase was considered 
in \cite{lutz,shura4}. In order to model the situation when the external 
signal is close to periodic one, but has a finite width of the spectral line 
Neiman and Schimansky-Geier  \cite{shura4} proposed to consider 
the harmonic noise as the 
input signal. Using the cumulant analysis and computer simulations they 
showed that the effect of SR takes place for harmonic noise as well and
the width of the spectral line of the input signal at the output power 
spectrum can be decreased via SR.

Harmonic noise $y(t)$ is defined by the following 
two-dimensional SDE \cite{lutz,shura4}:
\begin{equation}
\dot{y}\: =\: s,\hskip 0.5cm \dot{s}\: =\: -\Gamma s -\Omega^2 y + 
\sqrt{2\varepsilon \Gamma}\: \zeta (t),
\label{harmn}
\end{equation}
where $\zeta (t)$ is the zero-mean Gaussian noise with 
$\mean{\zeta (t)\: \zeta (t')}\: =\: \delta (t-t')$.
It is necessary to note, that Gaussian noise $\zeta(t)$ is statistically 
independent from the noise $\xi(t)$ in (\ref{model}).
Equation (\ref{harmn}) determines the two-dimensional Ornstein-Uhlenbek 
process $y(t),\: s(t)$ with the power spectrum 
\begin{equation}
S_{yy}(\omega)\: =\: \frac{\Gamma \varepsilon}
{\omega^2\: \Gamma^2 + (\omega^2 - \Omega^2)^2},
\label{hspectr}
\end{equation}
which for $\Omega^2>\Gamma^2 /4$ has a peak at the frequency 
$\omega_p \: =\: \sqrt{\Omega^2 - \Gamma^2 /2}$ with the width 
\begin{equation}
\Delta \Omega \: =\: \sqrt{\omega_p^2+\Gamma \omega_1}-
\sqrt{\omega_p^2-\Gamma \omega_1},
\end{equation}
where $\omega_1 \: =\: \sqrt{\Omega^2-\Gamma^2/4}$. 
The mean square displacements $\mean{y^2}=\varepsilon/\Omega^2, 
\mean{s^2}=\varepsilon, \mean{ys}=0$.
The increase of the parameter $\Gamma$ causes the widening of the spectral 
line  \cite{shura4}.
We used harmonic noise as the input signal in (\ref{model}) to carry out the 
multifractal analysis of SR in the case of  aperiodic driving force.
As in the previous subsection, we analyzed the return time sequences 
containing the same amount of points as before and normalized on the 
period $T=2\pi/\Omega$.
To compare our results with those obtained previously,
we choose the same values of parameters for numerical simulations as 
in \cite{shura4}.
The results of our computations are presented in Fig.~7.
The difference between the maximal and minimal H\"older exponents 
takes its minimal value for an optimal noise intensity as in 
the case of periodic driving. The obtained results are in good agreement with 
the results of \cite{shura4}. 
\begin{figure}
\epsfig{figure=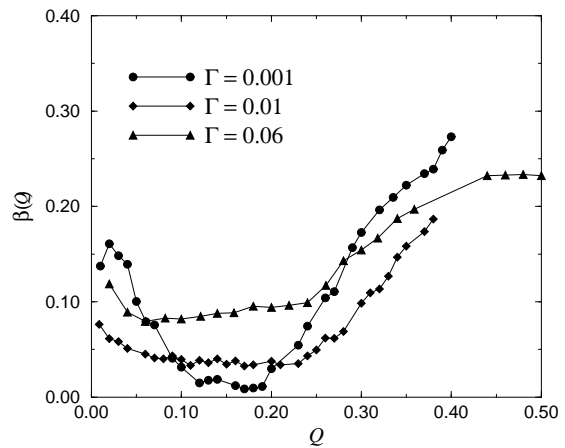,width=7.5cm}
\vskip 0.2cm
\begin{minipage}{8.2cm}
\caption{The degree of multifractality of the bistable system (\ref{model}) 
response to the harmonic noise vs. noise intensity for the different 
values of the dissipation parameter in (\ref{harmn}). 
Other parameters are $\Omega\: =\: 0.1, \varepsilon \: =\: 0.025$.}
\label{harmonica}
\end{minipage}
\end{figure}

\noindent
The degree of multifractality behaves as the 
relative width of the  output  spectral line used in \cite{shura4}
demonstrating the minimum at a close value of the noise intensity.
The decrease of $\Gamma$ in (\ref{harmn}) leads to the regularization of the 
input signal that makes  the minimum in $\beta (D)$ more 
pronounced. 

\subsection{Chaotic input signal}
Now, let us take as the input signal in (\ref{model}) the slowly varying 
subthreshold chaotic signal generated by the Lorenz system which is governed 
by the following ordinary differential equations:
\begin{eqnarray}
\label{lorenz}
\nonumber
\dot{y_1} &=& 10\: (y_2 - y_1)\nu ,\\ 
\dot{y_2} &=& (28\: y_1 - y_2-y_{1}y_{3})\nu ,\\ \nonumber
\dot{y_3} &=& (y_{1}y_2-8/3\: y_3)\nu,
\end{eqnarray}
where $\nu$ is the small rationing constant slowing chaotic oscillations.
Lorenz attractor existing in the phase space of this system has a 
thin multifractal structure \cite{wiklund}. 
\begin{figure}
\centering{\epsfig{figure=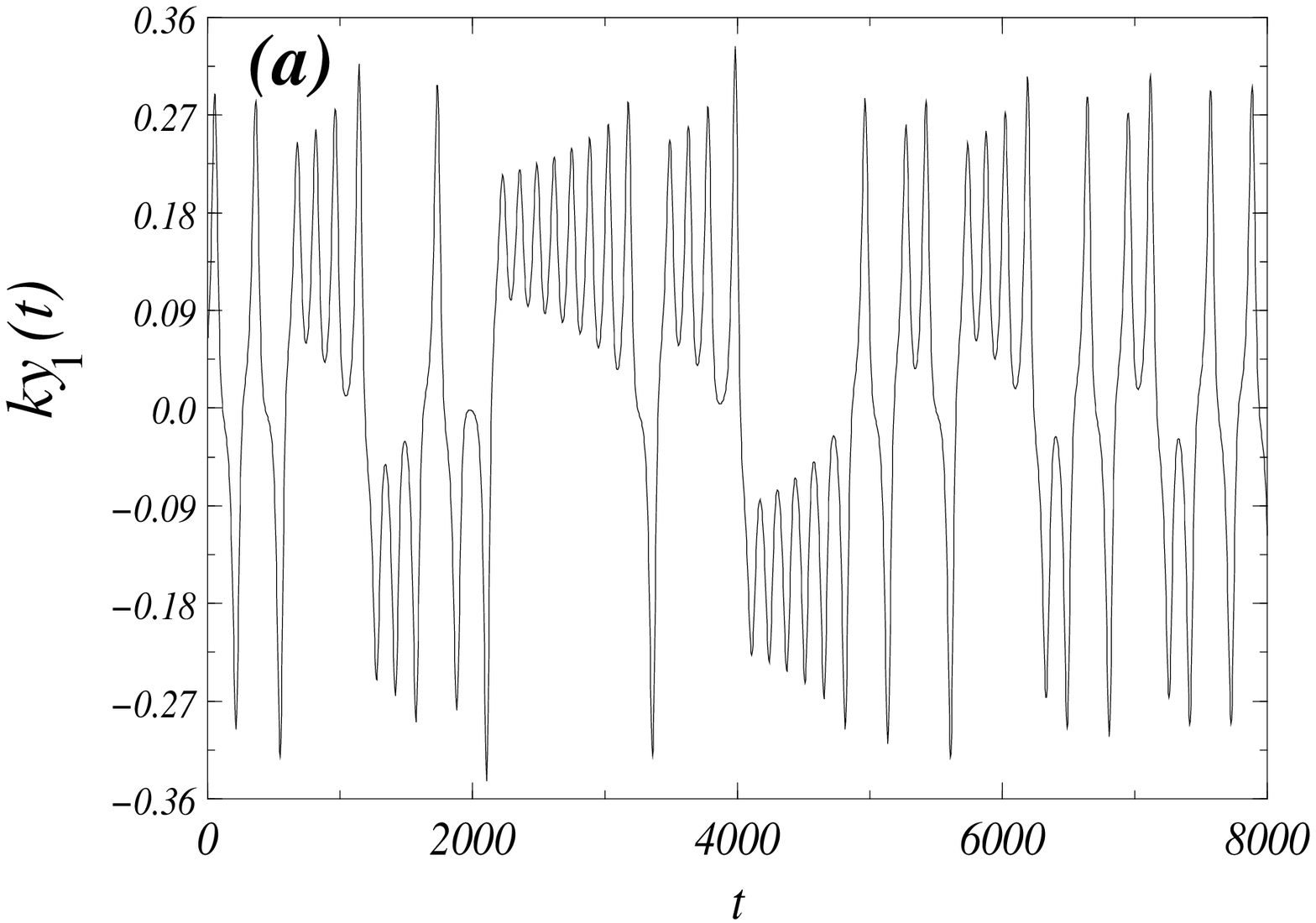,width=8cm,height=6cm}}
\vskip -0.9cm
\centering{\epsfig{figure=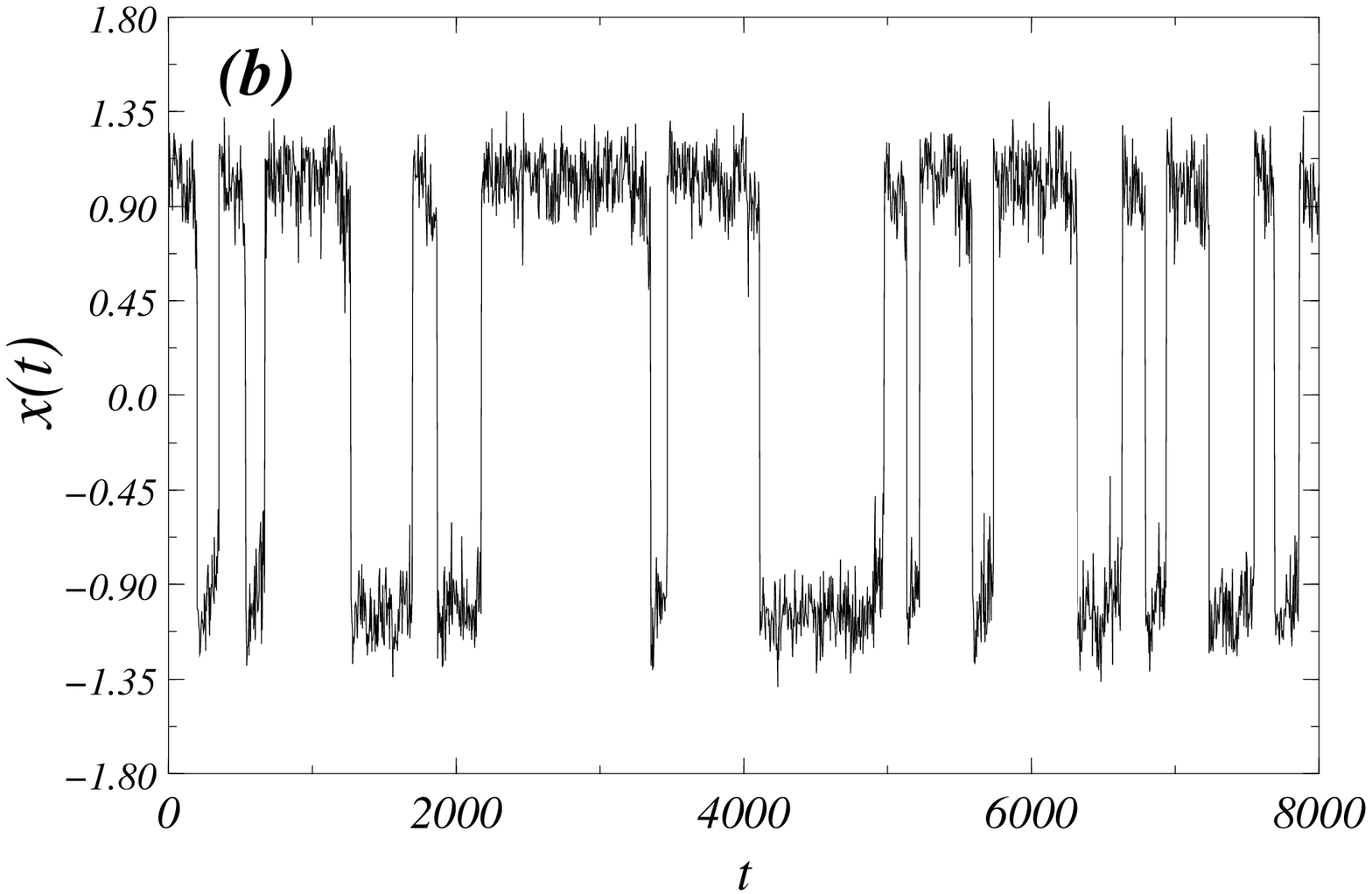,width=8cm}}
\vskip 0.2cm
\begin{minipage}{8.2cm}
\caption{The input (a) and output (b) signals in the regime of stochastic 
synchronization for the case of chaotic driving signal generated by the 
Lorenz signal. The parameters are  $k=0.0188$, $Q=0.03$, $\nu =0.005$.}
\end{minipage}
\end{figure}

\begin{center}
\begin{figure}
%%\vskip 0.2cm
\epsfig{figure=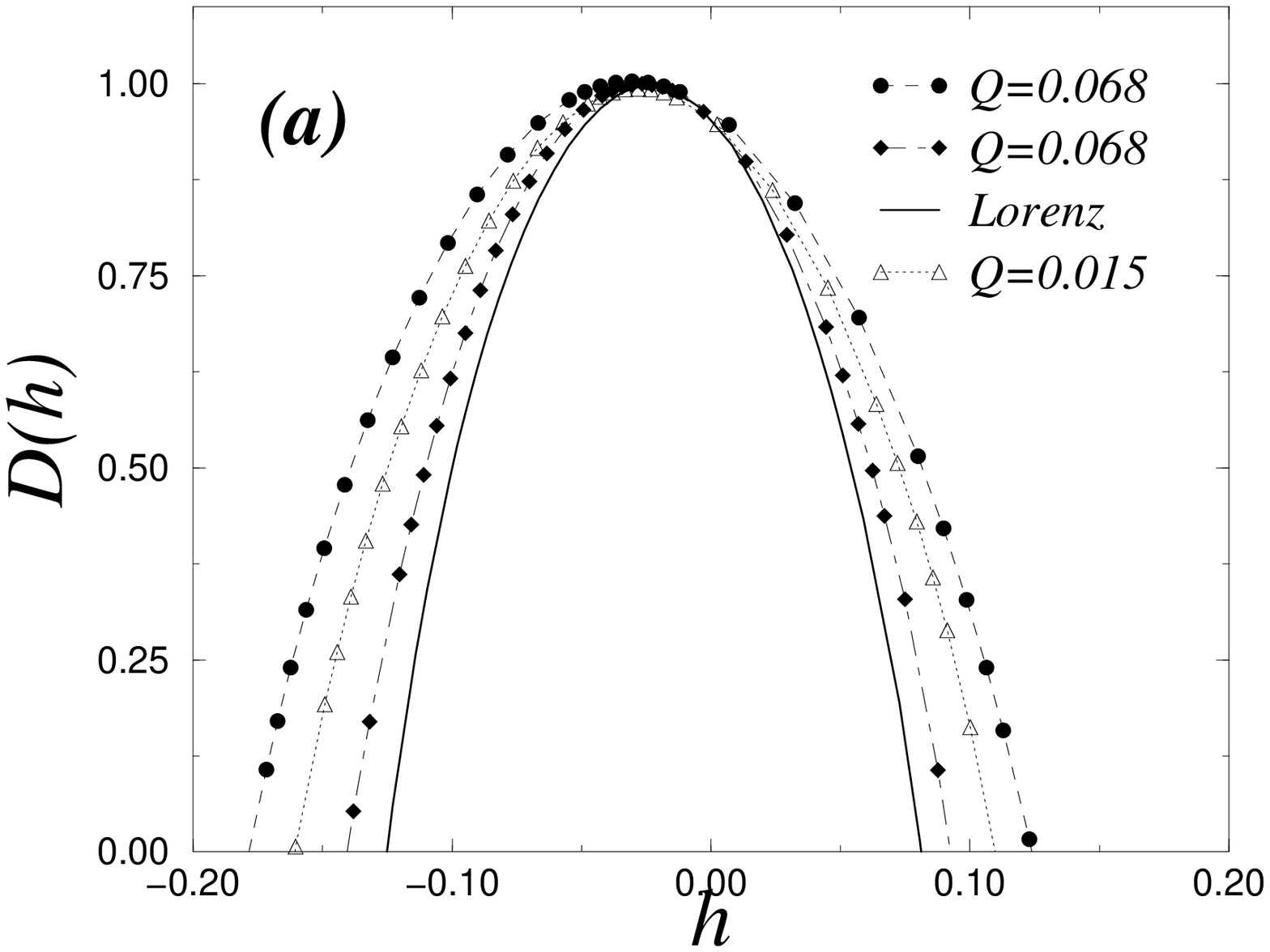,width=6.cm,height=3.8cm}\\
\hskip 1mm
\epsfig{figure=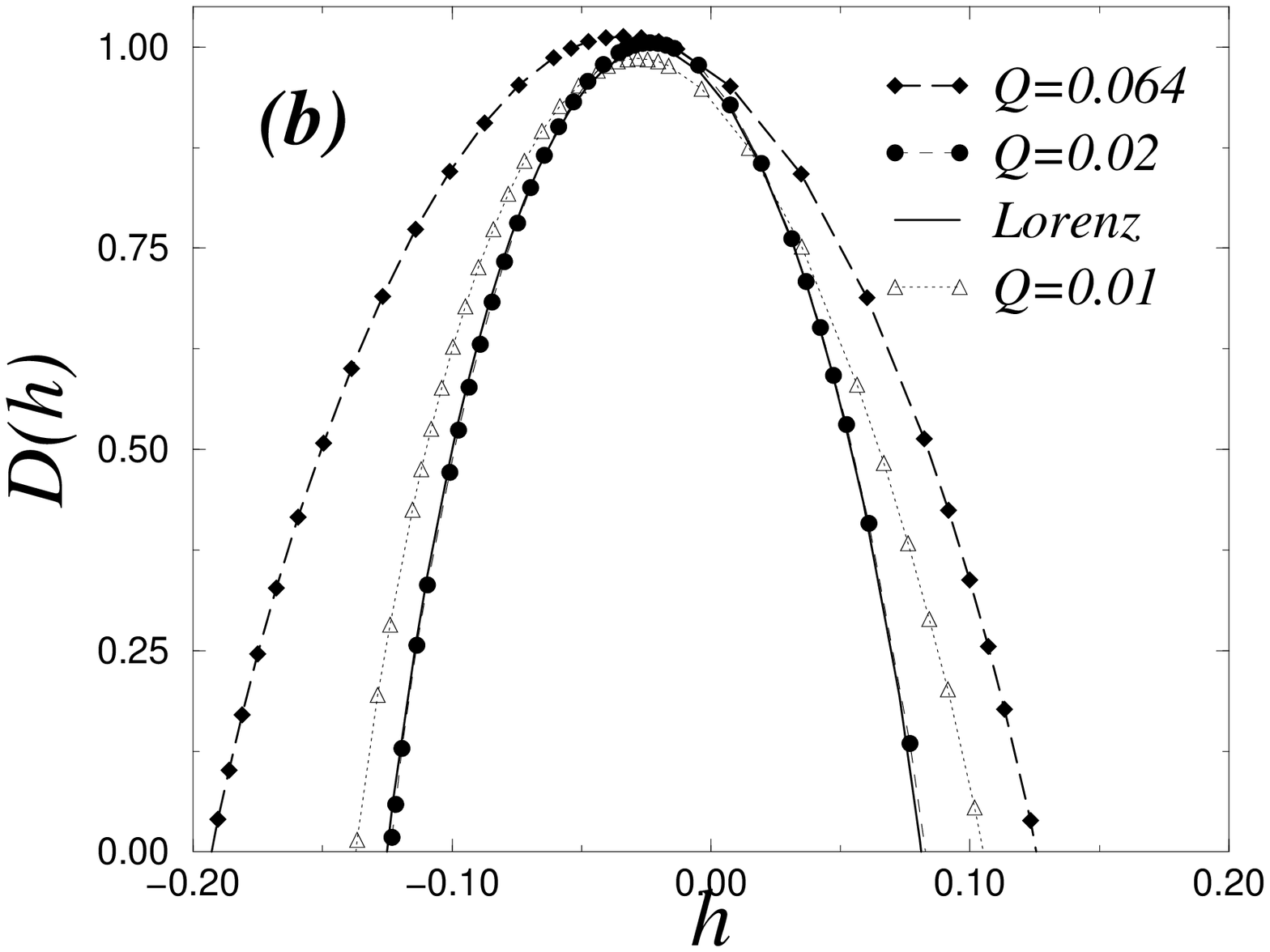,width=6.cm,height=3.8cm}\\
\vskip 0.2cm
\begin{minipage}{8.2cm}
\caption{Singularity spectrums for different values of noise intensity $Q$
and of the rationing constant: (a) $k=0.01$, (b) $k=0.0188$}
\end{minipage}
\end{figure}
\end{center}

\begin{figure}
\centering{\epsfig{figure=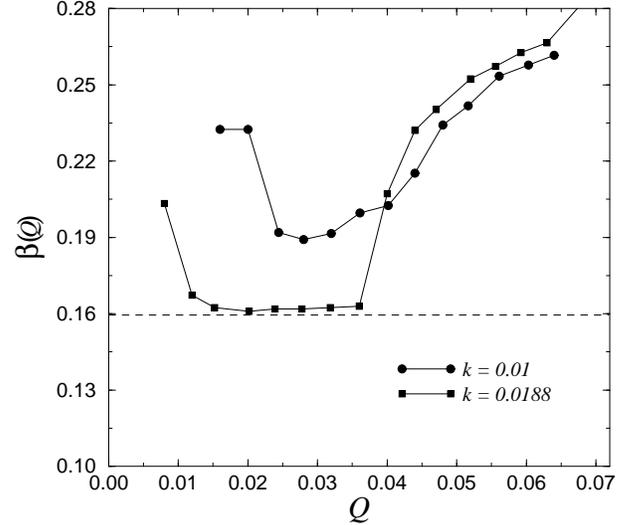,width=8.2cm}}
\begin{minipage}{8.3cm}
\caption{The degree of multifractality vs. noise intensity for the different 
values of chaotic signal amplitude. The width of the singularity spectrum 
of the input signal is represented by the dashed line. 
The Gaussian function was used
as the analyzing wavelet.}
\end{minipage}
\end{figure}
\noindent
The power spectrum of chaotic 
oscillations calculated for $y_{1,2}$-variables does not contain any sharp 
peaks in this case. 
The input signal  has a form of random  process of switchings 
between two metastable states (see Fig.~8 $(a)$).  
It is reasonable to calculate the return time sequences for the input and 
output signals
and then to try to estimate the distortion of the signal in stochastic 
resonator using singularity spectrum.
We used the variable $k\cdot y_1 (t)$ from the 
system of Eq. (\ref{lorenz}) as the input 
signal in our simulations, here $k$ is a small positive rationing constant.
As follows from the results presented in Fig.~9 $(a)$, 
singularity spectrums of the input and output signals become more close 
to each other for some optimal noise level. 
Moreover, for some values of the parameter $k$ the noise-enhanced phase 
coherence between chaotic signal and response is observed \cite{me}.
In this case, the switchings in the input signal are in phase  
with the switchings in stochastic bistable system (see Fig.~8).
Singularity spectrums of the input signal and of the response are coincide 
(see Fig.~9 $(b)$) in the regime of phase locking that means 
the passing of chaotic signal 
through the stochastic resonator without any distortions.
This is also illustrated by the dependence of the singularity spectrum's width 
on the noise intensity for  bistable system response  that is presented 
in Fig.~10.
For some optimal noise intensity, $\beta$ takes its minimal value closing to 
the value corresponding to the width of the input signal's singularity 
spectrum.
As was mentioned above, the input and output singularity spectrums coincide  
in the regime of stochastic synchronization  that causes the coincidence of 
$\beta$ with the multifractality degree of the external chaotic signal in some 
finite range of noise intensities.  
\section{multifractal analysis of neuron spike trains}
One of the reason for unremitting interest in SR is  the possibility to 
model on its base different cooperative effects and the  process of the 
information transfer in various biological sensory systems operating in  
natural noisy environment.
At present, there are a lot of experimental results showing that sensory 
neurons of different live organisms are able to demonstrate SR 
\cite{biosr,anishchenko}.
In order to estimate the enhancement of a response, the signal-to-noise ratio 
or different cross-correlation measures are  usually used \cite{collins,biosr}.
It is very interesting to use the above multifractal approach to analyze 
the spike trains generated by a stochastic neuronal model.

We took as a model the Fitzhugh-Nagumo system \cite{hohen} operating in 
 excitable regime and driven by a mixture of the internal noise and a 
subthreshold  stochastic spike train generated by another similar system 
detuned from the first one on a control parameter. 
The unidirectionally coupled  neuron systems are described by the following 
stochastic differential equations:
\begin{eqnarray}
\label{neurons}
\nonumber
\mu \dot{x_1} &=& x_1 -\frac{x_{1}^{3}}{3}-y_1 \\ 
\dot{y_1} &=& x_1 + a_1 +k\, x_2 +\sqrt{2\, Q_1}\: \xi_1 (t)\\ \nonumber
\mu \dot{x_2} &=& x_2 -\frac{x_{2}^{3}}{3}-y_2 \\ \nonumber
\dot{y_2} &=& x_2 + a_2  +\sqrt{2\, Q_2}\: \xi_2 (t) \nonumber
\end{eqnarray}
where $a_{1}$ ($Q_1$) and $a_{2}$ ($Q_2$) are, respectively, the 
control parameters (noise intensities) of the 
subsystems $(x_1,y_1)$ and $(x_2,y_2)$, $\xi_{1}$ and $\xi_{2}$ are 
the statistically independent Gaussian white noise with 
the zero mean, $k$ is a small rationing constant as before and 
$\mu \ll 1$ is a small parameter allowing one to separate all 
motions in the fast and slow ones.
The values of control parameters and noise intensities in 
subsystems are different and varied independently from each other.
Thus, we can  consider the system of Eq. (\ref{neurons}) as 
a model of a single neuron embedded in a network and driven by
both the internal noise and summed output of the neibouring neurons 
that can be modeled as a stochastic spike train. 
Interspike intervals (ISI) widely used in neuroscience as the typical neuron 
signals will play the role of signals under study in our consideration. 
The input stochastic spike train generated by the second neuron 
is characterized  by a continuous singularity spectrum having a finite width 
as well as the  chaotic input signal from the Lorenz system. 
\begin{figure}
\centering{\epsfig{figure=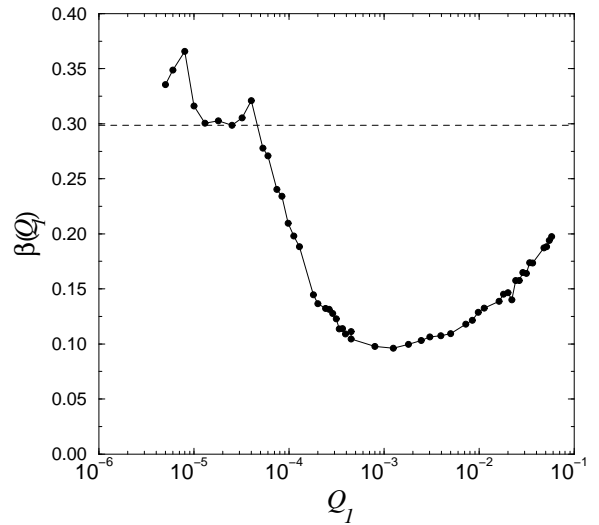,width=8cm}}
\begin{minipage}{8.2cm}
\caption{Degree of multifractality of the first neuron response in 
Eq. (\ref{neurons}) vs internal noise intensity $Q_1$. 
Parameters: $ \mu =0.01,\: a_1 =1.05,
\:  a_2 =1.07,\: Q_2 = 0.02,\:k=0.025$. The width of the input signal 
singularity spectrum is labeled by the dashed line.
The Gaussian function was used as the analyzing wavelet.}
\label{neurobeta}
\end{minipage}
\end{figure}
The calculated widths of the singularity spectrum for the ISI 
generated by the first neuron are shown in Fig.~11. 
It can be seen that the dependence of $\beta$ on the internal noise 
intensity is characterized by two different minimums corresponding to  
two different effects taking place in the system (\ref{neurons}) for small 
and large internal noise level, respectively.
The first minimum appearing at the comparatively small noise intensity 
corresponds to the effect of aperiodic stochastic resonance when a 
weak internal noise enhances the response of the neuron model optimizing 
the transmission of the input signal. As clearly seen from 
Fig.~11, the width of 
singularity spectrum calculated on the first neuron ISI is very close to 
the input one for some optimal values of noise level. Further increase of 
noise intensity makes singularity spectrum of response more narrow that can 
be considered as a manifestation of stochastic resonance without input
signal \cite{haken} called coherence resonance in \cite{pikovsky}. 
Indeed, in the case of sufficiently 
large noise intensity neuron already cannot  distinguish the structure of 
 the noisy input signal. It operates as an oscillator whose time scale
is controlled by noise. For some optimal value of the internal noise 
intensity $Q_1$ oscillations of the first neuron become close to the 
periodic one that is the essence of coherence resonance \cite{pikovsky}.
At the moment when coherence resonance is observed $\beta$ takes its second 
deeper minimum reflecting the noise-enhanced ordering of ISI. 
\section{conclusions}
We have studied the phenomenon of stochastic resonance in terms of 
the multifractal formalism revisited with wavelets \cite{muzy3}. 
We observed that for some optimal noise intensity the degree of 
multifractality of the response, defined as a width of the singularity 
spectrum, takes its minimal value. Moreover, the qualitative change of 
its structure takes place in the regime of stochastic synchronization. 
In the region of the noise-enhanced phase locking it shrinks to the single 
point with zero H\"older exponent.
We have shown that the width of the singularity spectrum calculated over the 
return time sequence can be effectively used as the measure 
characterizing the response of a noisy nonlinear system in a wide range of 
the driving amplitudes and frequencies. As follows from our 
numerical results this measure can be successfully used both for periodic and 
aperiodic (stochastic or chaotic) driving signals. 
Moreover, it has allowed us to estimate the degree of coherence for the 
unidirectionally coupled stochastic neurons model operating in excitable 
regime. By using the introduced  
measure, we successfully diagnose both aperiodic stochastic resonance and 
coherence resonance which take place in the model under study for the small 
and large noise intensities, respectively. 

The proposed approach has a number of benefits in comparison with 
the traditionally used measures such as SNR, SPA, residence time distributions,
coherence function and others. 
These measures use the averaging procedure for their calculation that 
leads to the loss of information about nonlinear interaction between 
Fourier phases in response. This information is very important both for deeper 
understanding of the essence of SR and for more sensitive diagnostic of 
SR in full-scale experiments. The multifractal formalism based on wavelet 
calculations allows to study the temporal structure of the response. It  
catches all even weak non-stationarityies in a return times sequence under study 
that makes it a very powerful tool for diagnostic of SR and stochastic 
synchronization. The introduced measure demonstrates the behavior which 
in a very good agreement with the behavior of traditional quantitative 
characteristics of SR. It is universal in relation to the kind of the input 
signal and able to catch noise-induced effects even from very short time 
series. Later has the especial importance for the analysis of real signals.

The presented approach to the study of scaling features of motion in 
stochastic systems may be very fruitful also in the case of the Brownian 
motion in periodic potential under the action of random forces.
It will be the task of our future investigations.

\acknowledgments 

All the computations of singularity spectrums in this paper have been made 
using free GNU licensed software LastWave \cite{lastwave}.
This work was supported in part by the National Science Council of the 
Republic of China (Taiwan) under Contract No. NSC 89-2112-M-001-005.

\end{multicols}

\newpage
\vskip 1cm
\centerline{FIGURE CAPTIONS:}
\vskip 1cm

Fig.~1. (a) -- the path of a Brownian particle; (b) -- the modulus-maxima 
skeleton of the random signal pictured in (a); (c) -- the dependence 
$\tau (q)$; (d) -- the  singularity spectrum. 
The first derivative of the Gaussian function was used as the analyzing wavelet.
\vskip 0.5cm

Fig.~2. (a) $D(h)$ in the case of the weak periodic 
input signal $A\: =\: 0.081, \Omega\: =\: 0.004$ for different values of 
noise intensity $Q$; 
(b) $\tau(q)$ corresponding to the singularity spectrums, $D(h)$, 
in (a) for different values of noise intensity; 
the sequences of the return times normalized on the external force period and 
corresponding singularity spectrums in the case of $A\: =\: 0.286, 
\Omega\: =\: 0.004$ for $Q=0.040$ ((c) and (d)) and $Q=0.012$ ((e) and (f));
The Gaussian function was used as the  analyzing wavelet.
\vskip 0.5cm

Fig.~3. Degree of multifractality vs. noise intensity
for different values of the periodic force
amplitude: $A\: =\: 0$ (stars), $A\: =\: 0.081$
(triangles), $A\: =\: 0.152$ (diamonds), $A\: =\: 0.268$ (circles).
The frequency of external force has the same value as in Fig.~2.
The Gaussian function was used as the  analyzing wavelet.
\vskip 0.5cm

Fig.~4. Regions of stochastic synchronization constructed by means of 
effective diffusion constant (dashed line) and of multifractality degree
(solid line). Driving frequency is $\Omega \:=\: 0.004$. The Gaussian
function was used as the analyzing wavelet.
\vskip 0.5cm 

Fig.~5. Degree of multifractality vs. noise intensity calculated for 
indicated values of the length of the return time sequence. 
The amplitude and frequency of the driving force are 
$A\: =\: 0.12$, $\Omega\: =\: 0.004$.
\vskip 0.5cm

Fig.~6. Spectral power amplification vs. noise intensity for different
values of the driving amplitude. Driving frequency is $\Omega \:=\: 0.004$.
\vskip 0.5cm

Fig.~7. Degree of multifractality of the bistable system (\ref{model})
response to the harmonic noise vs. noise intensity for the different
values of the dissipation parameter in (\ref{harmn}).
Other parameters are $\Omega\: =\: 0.1, \varepsilon \: =\: 0.025$.
\vskip 0.5cm

Fig.~8. The input (a) and output (b) signals in the regime of stochastic
synchronization for the case of chaotic driving signal generated by the
Lorenz signal. The parameters are  $k=0.0188$, $Q=0.03$, $\nu =0.005$.
\vskip 0.5cm

Fig.~9. Singularity spectrums for different values of noise intensity $Q$
and of the rationing constant: (a) $k=0.01$, (b) $k=0.0188$.
The Gaussian function was used as the analyzing wavelet.
\vskip 0.5cm

Fig.~10. Degree of multifractality vs. noise intensity for the different
values of chaotic signal amplitude. The width of the singularity spectrum
of the input signal is represented by the dashed line.
The Gaussian function was used as the analyzing wavelet.
\vskip 0.5cm

Fig.~11. Degree of multifractality of the first neuron response in
Eq. (\ref{neurons}) vs internal noise intensity $Q_1$.
Parameters: $ \mu =0.01,\: a_1 =1.05,
\:  a_2 =1.07,\: Q_2 = 0.02,\:k=0.025$. The width of the input signal
singularity spectrum is labeled by the dashed line.
The Gaussian function was used as the analyzing wavelet.

\end{document}